\documentclass{appolb}
\usepackage{graphicx}
\usepackage{epsfig,amsmath}
\usepackage{amsfonts}
\usepackage{amsmath,epsfig}
\usepackage{graphicx}
\usepackage{dcolumn}
\usepackage{bm}
\usepackage{amssymb}
\usepackage{amsmath}%

\begin{document}
\title{Dense hadronic matter in neutron stars%
\thanks{Presented at Excited QCD 2014, Bjelasnica Mountain, Sarajevo}%
}
\author{Giuseppe Pagliara$^a$, Alessandro Drago$^a$, Andrea Lavagno$^b$ and Daniele Pigato$^b$
\address{$^a$Dip.~di Fisica e Scienze della Terra dell'Universit\`a di Ferrara and INFN
Sez.~di
Ferrara, Via Saragat 1, I-44122 Ferrara, Italy}
\address{$^b$Department of Applied Science and Technology, Politecnico
di Torino and
Istituto Nazionale di Fisica Nucleare (INFN), Sezione di Torino, Italy}}

\maketitle
\begin{abstract}
The existence of stars with masses up to $2 M_{\odot}$ and the hints of the existence of
stars with radii smaller than $\sim 11$ km seem to require, at the
same time, a stiff and a soft hadronic equation of state at large
densities.  We argue that these two apparently contradicting
constraints are actually an indication of the existence of two
families of compact stars: hadronic stars which could be very compact
and quark stars which could be very massive. In this respect, a
crucial role is played, in the hadronic equation of state, by the
delta isobars whose early appearance shifts to large densities the
formation of hyperons. We also discuss how recent experimental
information on the symmetry energy of nuclear matter at saturation
indicate, indirectly, an early appearance of delta isobars in neutron
star matter.
\end{abstract}
  
\section{Introduction}
The discoveries of massive neutron stars, with $M=2M_{\odot}$, 
\cite{Demorest:2010bx,Antoniadis:2013pzd} represent a challenge for
nuclear and hadron physics: the central densities of these stellar
objects are in the range from three to seven times nuclear saturation
density, depending on the model adopted for calculating the nucleonic
equation of state, see \cite{Baldo:2013ska}. At such large densities,
new hadrons are likely to form, such as hyperons and delta isobars,
which however strongly soften the equation of state leading to a
maximum mass smaller than the measured masses. The softening of the
equation of state allows however to obtain stellar configurations
which can be very compact and thus compatible with the results of recent
analyses of the thermal emission of quiescent low-mass X-ray binaries
suggesting the existence of stars with radii smaller than $\sim 11$ km
\cite{Guillot:2013wu,Lattimer:2013hma}. Although these
analyses are still under debate, we investigate here what would they imply
for the composition of matter at high densities.

\begin{figure}[ptb]
\vskip 0.5cm
\begin{centering}
\epsfig{file=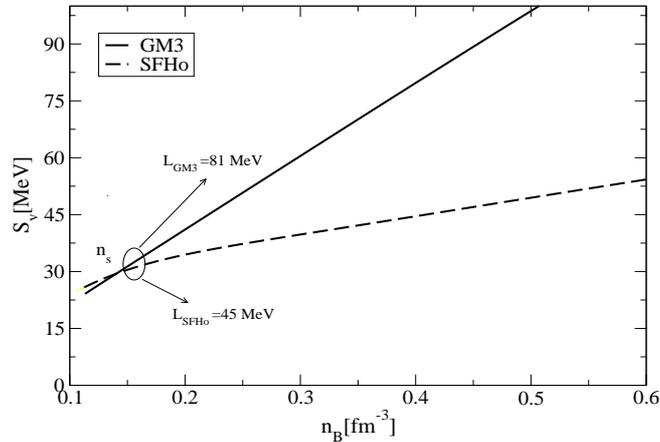,height=8.5cm,width=5.8cm,angle=-90}
\caption{Symmetry energy as a function of the baryon density: comparison between the GM3 equation 
of state \cite{Glendenning:1991es} and the recent SFHo equation of state 
\cite{Steiner:2012rk}.}
\end{centering}
\end{figure}
 
Presently, none of the proposed equations of state for dense matter
allows to fulfill at the same time the astrophysical constraints,
i.e. maximum mass of at least $2 M_{\odot}$ and radii $\lesssim 11$
km, and the hadronic physics constraints of the appearance, at large
baryon chemical potentials, of new degrees of freedom of the
baryon octet and decuplet. In Ref. \cite{Drago:2013fsa}, we argue
that a possible way out to this problem is that actually two families
of compact stars exist: hadronic stars which can be very compact (radii could be smaller than $\sim 10-11$ km) and
have maximum masses up to $\sim 1.5-1.6 M_{\odot}$ and quark stars which have larger radii and can
reach masses up to $2.75 M_{\odot}$, as resulting from pQCD
calculations \cite{Kurkela:2009gj}. For this scenario to be feasible,
the formation in the stellar matter of delta isobars is crucial and, as we will show in the
following, the recent constraints on the symmetry energy of nuclear
matter at saturation favor an early appearance of delta isobars.

\begin{figure}[ptb]
\vskip 0.5cm
\begin{centering}
\epsfig{file=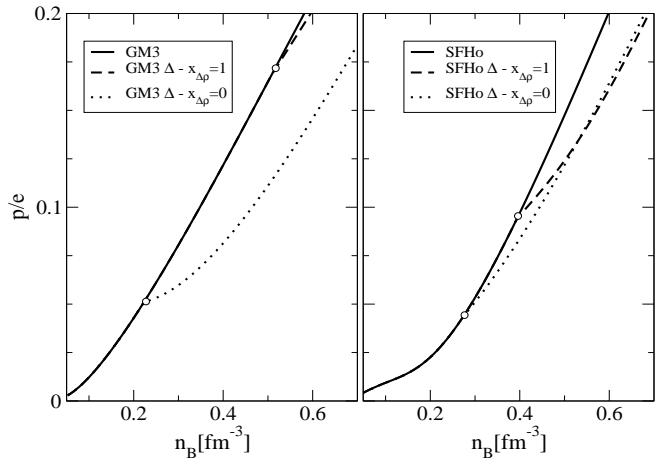,height=8.5cm,width=6cm,angle=-90}
\caption{Ratio between pressure and energy density as a function of the baryon density for GM3 and SFHo models 
(left and right panels respectively). Two cases for the coupling between $\Delta$ and $\rho$ are considered: $x_{\Delta\rho}=0,1$.  }
\end{centering}
\end{figure}

\section{Equation of state and mass-radius relations}
We adopt a Walecka-type relativistic mean field model for the hadronic
equation of state introduced in Ref.\cite{Steiner:2004fi}. In this
model, additional non-linear terms are added (in the vector
mesons sector) to the original Glendenning model
\cite{Glendenning:1984jr} which allow to better constrain the equation of
state at saturation by use of new experimental information on symmetry
energy $S_v$, giant monopole resonances and finite nuclei properties.  
In particular, we use the recent parametrization proposed in
\cite{Steiner:2012rk}, SFHo, but including also delta isobars and
hyperons. 
Let us first discuss the results obtained for $S_v$ as a
function of the baryon density $n_B$. In Fig. 1 we show $S_v$ for
the GM3 \cite{Glendenning:1991es} and the SFHo
models. Notice the splitting of the two results as the density exceeds the saturation density, 
with the SFHo result lying below the GM3 result.  
We remark that in the GM3 model no constraint is imposed in particular on the derivative with
respect to density of the symmetry energy at saturation, the parameter
$L$ \cite{Lattimer:2012xj} and which turns out to be of
about $81$MeV. On the other hand, in the SFHo model, the additional
parameters introduced in the Lagrangian, allow to fix $L$ to $\sim 45$ MeV,
a value compatible with the analyses of Ref.\cite{Lattimer:2012xj}, where a window of values of $L$ between $40$
and $60$ MeV has been obtained by use of laboratory and astrophysical
constraints. The term of the symmetry energy related to the interaction, as obtained
in the SFHo model, reads \cite{Steiner:2004fi}:
$\frac{g^2_{\rho}/m^2_{\rho}n_B}{8 (1+2g^2_{\rho}/m^2_{\rho}f)}$ where
$g_{\rho}$ and $m_{\rho}$ are the baryon-$\rho$ meson coupling and the
mass of the $\rho$ meson respectively and $f$ is a polynomial function
of the $\sigma$ and $\omega$ fields. In the GM3 model, $f=0$ and
$S_v$ increases linearly with the density. A more complicated dependence on the density arises 
in the SFHo model which however can be mapped into a GM3-like model by use of a density dependent coupling 
$g_{\rho}(n_B)$ which decreases as a function of the density. 
This parameter is crucial for computing the
thresholds of appearance of the different baryons:
depending on its value, delta isobars could appear after or before 
the hyperons as the density increases.
As discussed in \cite{Glendenning:1984jr}, among
the four isobars, the $\Delta^-$ is likely to appear first because it
is ``electric charge favored'' (the $\Delta^0$ chemical potential does not get a contribution from the electric charge
chemical potential and $\Delta^+$, $\Delta^{++}$ are electric charge
unfavored). However it is ``isospin unfavoured'' due to its isospin charge
$t_3=-3/2$. The coupling with the $\rho$ meson thus
affects more the threshold of the  $\Delta^-$ rather than
the thresholds of the hyperons.
In the calculations of Ref. \cite{Glendenning:1984jr} delta isobars appear after the hyperons and at
densities which are too high to be reached in compact stars. Of course
the crucial inputs for calculating the thresholds are the baryon-meson couplings
expressed as the ratios with the nucleon-meson couplings: $x_{i\sigma}=g_{i\sigma}/g_{N\sigma}$,
$x_{i\omega}=g_{i\omega}/g_{N\omega}$, $x_{i\rho}=g_{i\rho}/g_{N\rho}$
where $i$ runs over the hyperons and the delta isobars. For
calculating the beta stable equation of state needed for compact
stars, the couplings of the hyperons are fixed as in
\cite{Drago:2013fsa} while for the delta isobars we set:
$x_{\Delta\omega}=x_{\Delta\rho}=1$ and $x_{\Delta\sigma}$ is varied
in the interval $1-1.15$.
In Fig. 2 we display the ratio between pressure and energy density
(which provides a measurement of the stiffness of the equation of
state) for the GM3 and SFHo models with $x_{\Delta\omega}=x_{\Delta\rho}=x_{\Delta\sigma}=1$. For the sake of discussion also the case $x_{\Delta\rho}=0$ is included 
(here the hyperons degrees of freedom are
artificially switched off). Notice that for $x_{\Delta\rho}=1$, which
is the standard choice \cite{Glendenning:1984jr}, the delta isobars
appear at a density slightly above $0.5$ fm$^{-3}$ in GM3 and slightly
below $0.4$ fm$^{-3}$ in SFHo. In turn this implies that in the GM3
model hyperons appear before the delta isobars, as found in
\cite{Glendenning:1984jr}, shifting their threshold to very large
densities. On the other hand, in SFHo it is the opposite, delta
isobars appear first and they shift to large densities the hyperons. As
explained before this different behavior is due to the coupling
with the $\rho$ meson: while in the GM3 model this coupling is
constant, in the SFHo model, effectively, it decreases with the density thus favoring states, as the $\Delta^-$, with negative
isospin charge. This is also clear when looking at the curves obtained for $x_{\Delta\rho}=0$: in GM3 a strong reduction of the
threshold density is obtained (of about $0.3$ fm$^{-3}$ with respect to the case
 $x_{\Delta\rho}=1$ ) while in SFHo
it is reduced of only $0.1$ fm$^{-3}$.

In Fig. 3, we show the
mass-radius relations of compact stars, including pure nucleonic stars
(black line), hadronic stars with only delta isobars (green dashed
line), hadronic stars with hyperons and delta isobars (red
lines) and finally pure quark stars (blue line, same as in
\cite{Drago:2013fsa}). The stellar configuration at which 
the green dashed line and the black line separate has a central density corresponding to
the threshold for the formation of delta isobars. Similarly, for the formation of hyperons (continuous red line and green dashed line) 
which in the SFHo model appear after the delta isobars.
We also display the two solar mass limit and the recent interval of
radii indicated by the analyses of Refs. \cite{Guillot:2013wu}. The two solar mass limit can be
reached only by quark stars (nucleonic stars also reach the limit but
only if hyperons and delta degrees of freedom are artificially
switched off when computing the equation of state). On the other hand,
configurations with small radii and masses close to the canonical $1.4
M_{\odot}$ are obtained with the hadronic equation of state that
includes both hyperons and delta isobars (see also
\cite{Schurhoff:2010ph}) but only if the coupling of the delta with
the $\sigma$ meson is slightly larger than the coupling of the nucleon with the same meson, 
i.e. $x_{\Delta\sigma}=1.15$ (similar effects are obtained by reducing $x_{\Delta\omega}$ or $x_{\Delta\rho}$ ).
Arguments in favor of values of $x_{\Delta\sigma}$ larger than one can be
found in \cite{Jin:1994vw,Lavagno:2012bn}. As we have proposed in
\cite{Drago:2013fsa}, if small radii stars do really exist together with massive stars,
the scenario of coexistence of two
families of compact stars is strongly favored. In this scenario, most
of the stars are actually hadronic stars and only very massive stars
are composed by pure quark matter. The mechanism which allows to
populate the quark star branch and the observational consequences of
such a conversion process have been discussed in several papers
\cite{Drago:2004vu,Bombaci:2004mt,Drago:2005yj,Pagliara:2013tza,Drago:2013fsa,Buballa:2014jta}.
Notice that the early appearance of delta isobars is crucial for this
scenario to be viable: they indeed delay the appearance of hyperons
which, once formed, are responsible for the seeding of stable strange
quark matter and for the subsequent conversion process.
\begin{figure}[ptb]
\vskip 0.5cm
\begin{centering}
\epsfig{file=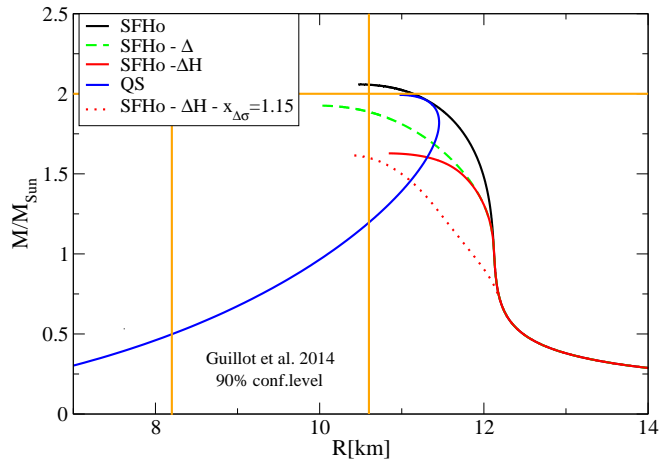,height=8.5cm,width=6cm,angle=-90}
\caption{Mass radius-relations for different equations of state together with the maximum mass constraint and the radii
window indicated by the analyses in \cite{Guillot:2013wu}.}
\end{centering}
\end{figure}
\section{Conclusions}
The new constraints on the symmetry energy at saturation, in
particular the $L$ parameter, seem to favor an early appearance of
delta isobars in dense matter. These degrees of freedom, together with
hyperons, must be included in every calculation aiming at
understating the structure of compact stars. The necessary softening
of the equation of state allow for the existence of very compact stars
although not very massive. However, the tension between the existence
of massive neutron stars (with candidates with masses even
larger then $2 M_{\odot}$ ) and the recent indications of existence of
very compact stars could be relieved within a scenario of coexistence
of two families of compact stars. In particular heavier stars are, in
our proposal, quark stars. These stellar objects, a part from their
masses and radii larger than the one of hadronic stars, should show
anomalous cooling histories and spinning frequency
distributions. Moreover, in basically all the processes of merger of
neutron stars we expect that the remnant, before collapsing to a
black hole, is a quark star.

\end{document}